\documentclass{article}

\usepackage[preprint,nonatbib]{confstyle}

\usepackage[utf8]{inputenc} 
\usepackage[T1]{fontenc}    
\usepackage{hyperref}       
\usepackage{url}            
\usepackage{booktabs}       
\usepackage{amsfonts}       
\usepackage{nicefrac}       
\usepackage{microtype}      
\usepackage{xcolor}         
\usepackage{graphicx}  
\usepackage{setspace}
\usepackage{multirow}
\usepackage{hyperref}

\usepackage{mathptmx} 
\usepackage{times} 
\usepackage{amssymb}  

\usepackage{listings} 

\usepackage{algorithm}  
\usepackage{algorithmic}
\usepackage{amsmath}

\usepackage{balance}
\usepackage{adjustbox}
\usepackage{tabularx}

\title{\Large Exploring Human-Machine Coexistence in Symmetrical Reality}

\author{Zhenliang Zhang\thanks{E-mail: zlzhang@bigai.ai}\\
State Key Laboratory of General Artificial Intelligence, BIGAI
}

\makeatletter
\def\thanks#1{\protected@xdef\@thanks{\@thanks
        \protect\footnotetext{#1}}}
\makeatother

\begin{document}

\maketitle

\begin{abstract}
In the context of the evolution of artificial intelligence (AI), the interaction between humans and AI entities has become increasingly salient, challenging the conventional human-centric paradigms of human-machine interaction. To address this challenge, it is imperative to reassess the relationship between AI entities and humans. Through considering both the virtual and physical worlds, we can construct a novel descriptive framework for a world where humans and machines coexist symbiotically. This paper will introduce a fresh research direction engendered for studying harmonious human-machine coexistence across physical and virtual worlds, which has been termed ``symmetrical reality''. We will elucidate its key characteristics, offering innovative research insight for renovating human-machine interaction paradigms.
\end{abstract}

\section{The Origin of Symmetrical Reality}

Humans and artificial intelligence (AI) entities act as the main agents interacting with their external environments. Historically, in the domain of human-machine interaction, humans have always been considered the principal agents, epitomizing a human-centric system of interaction. However, with the elevation of machine intelligence levels and the evolution of machine behaviors~\cite{rahwan2019machine}, machines are progressively exhibiting a high degree of autonomy, demonstrating human-like independence and autonomy during interactions with external environments. Particularly with the advent of large models such as ChatGPT, the immense potential of artificial intelligence is increasingly recognized by researchers. This suggests that as AI entities interact with humans, there will be a gradual shift away from the human-centric paradigm, establishing AI as another perceptual center of equal significance to humans. Such a transformation presents profound research value for the field of human-machine interaction.

Considering that artificial intelligence represents an entity not naturally present in the natural world and shares similar artificial characteristics with artificial virtual spaces, it can be posited that AI is originally an object within this artificial virtual space. Consequently, a symmetrical structure is formed between the artificial virtual world and the physical world. Specifically, while humans inhabit the physical space, AI entities reside in the virtual world and can have evolvable embodiments~\cite{gupta2021embodied}. Both humans and AI entities have the capacity to concurrently perceive the physical and virtual worlds. This establishes the symmetry regarding the perception and interaction processes between humans and AI, termed as ``symmetrical reality''~\cite{zhang2024emergence}.

\begin{figure}[tb]
    \centering
    \includegraphics[width=\linewidth]{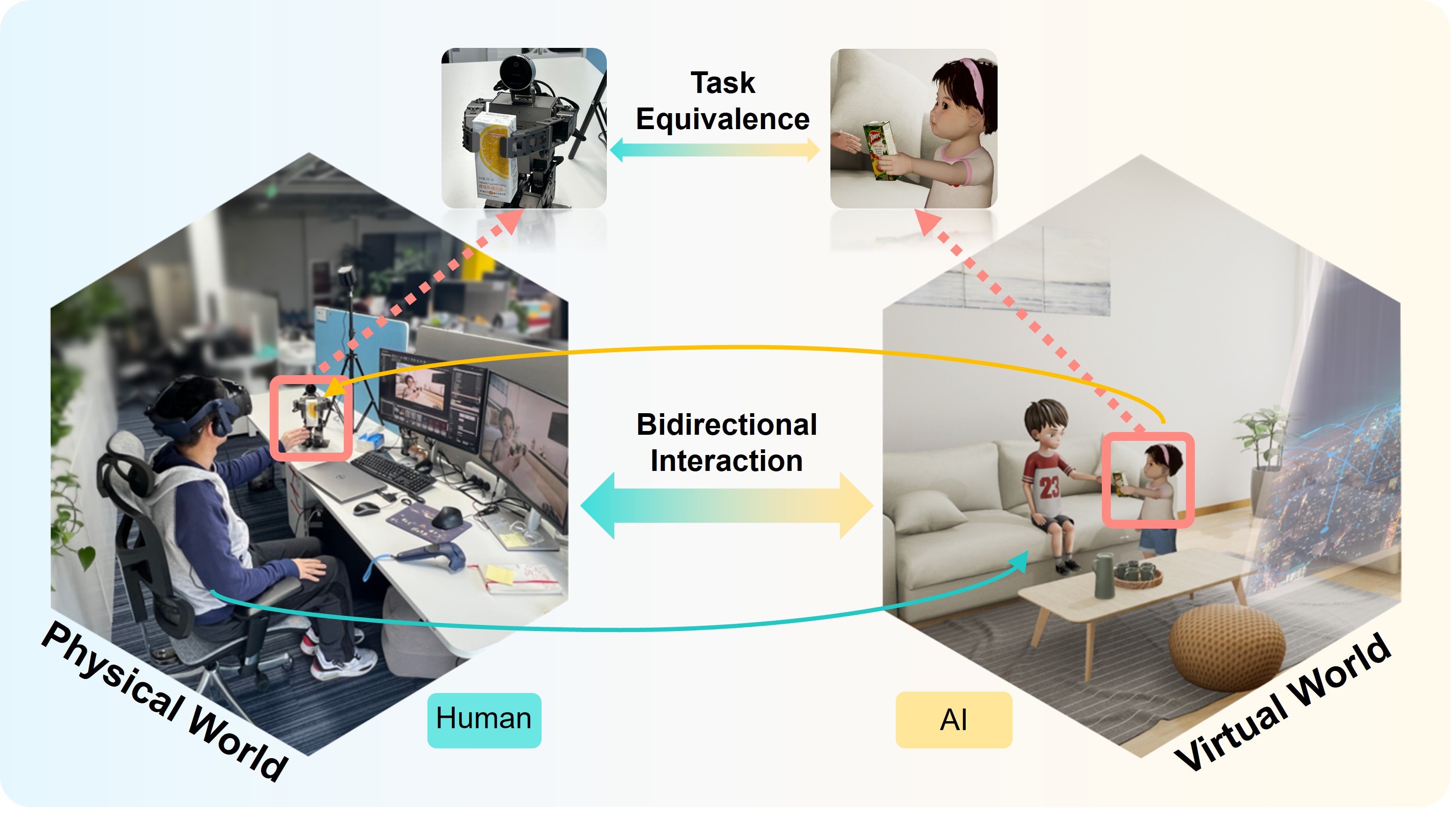}
    \caption{A conceptual illustration for a typical symmetrical reality system. The human user and the AI agent can interact with each other across the physical and virtual worlds. A task can happen in both worlds so that the physical and virtual worlds evolve simultaneously.} 
    \label{fig:firstpage}
\end{figure}

The concept of symmetrical reality is introduced to articulate the coexistence issue of humans and machines in a mixed reality environment, highlighting the symmetrical perception and interaction as the foundational form for future human-machine coexistence, as shown in Fig.~\ref{fig:firstpage}. Under this emerging scenario, the traditional human-centric paradigm of human-machine interaction should undergo significant modification, thus becoming competent to accommodate a dual-center system of humans and AI entities. Within such a framework, a new state of human-machine coexistence emerges, highlighting both the needs of humans (e.g., Maslow's hierarchy of needs~\cite{maslow1943theory}) and the needs of AI entities. The theory of needs for AI essentially reflects its inherent value system, which is pivotal for harmonious human-machine coexistence. The symmetrical reality framework offers a fresh avenue for discussing these critical issues.

\section{Why Symmetrical Reality Matters?}
Symmetrical reality emerges as a novel paradigmatic descriptor for human-machine interaction based on the rapid evolution of AI, and it also represents a fundamental solution proposed for the long-term harmonious coexistence between humans and machines. The symmetrical reality framework embodies a dual-center architecture between humans and AI entities, emphasizing the AI's significant role within this structure because of its autonomy and independence. The perceptual and interactional structure between humans and AI entities is symmetrical, which is of pivotal importance in a blended physical-virtual world. We shall now delve into the perception and interaction of both humans and AI entities.

Perceiving the external world is a fundamental capability of both humans and AI entities. However, simultaneous perception of both the physical and virtual worlds is a distinctive trait emphasized for human-machine coexistence. Within the symmetrical reality framework, both the physical and virtual worlds coexist concurrently; humans inhabit the physical world while AI entities reside in the virtual world. Humans can observe their physical surroundings, and AI entities can observe their virtual environment. Once communication mechanisms are established between the physical and virtual worlds, humans can access the virtual world, and AI entities can engage with the physical world. Consequently, both humans and AI entities can perceive both the physical and virtual realms, forming a symmetrical structure of perception. Here, symmetry denotes the equivalency in perceptual capabilities between humans and AI entities.

Interacting with the external environment is also a foundational ability for both humans and AI entities. Given the simultaneous perception of both virtual and physical worlds, interactions with the external environment become more intricate. Human interactions with the physical world are intuitive, and their engagements with the virtual realm (akin to playing VR games) are easily achievable. AI entities' operations within the virtual world can be considered foundational, given that virtual AI entities intuitively interacts with the virtual environment. The only non-intuitive aspect is the AI's interaction with the physical world, which requires embodiments like robotic arms. In summary, bidirectional human-machine interactions within a blended physical-virtual world introduce new challenges propelled by AI advancements, such as human-machine cooperation~\cite{crandall2018cooperating}. It is imperative to study how to utilize these technologies effectively while preventing potential misuse hazards. After all, the implications of AI entities' autonomy and independence on the external world remain an under-researched domain. Notably, this also marks a significant shift from traditional human-machine interaction paradigms.

\section{Autonomous Machine in Symmetrical Reality}

\subsection{Autonomous Machine and Key Features}
We argue that every autonomous machine should be equipped with value systems, which guides machines behaviors during the interaction between machines and humans. The so-called “value system” can generate the intrinsic motivation of AI machines, which makes a machine become truly autonomous. This kind of change weakens the boundary between machines and living organisms. Sometimes, the autonomous machines behave like humans in specific scenarios, when they have passed the Turing tests. We do not discuss about the machine consciousness due to its complexity, but the highly autonomous machine would affect human cognition and behaviors in many aspects.

According to the research about artificial general intelligence (AGI)~\cite{peng2024tong}, an intelligent machine should possess three features before it could reach AGI. 
The first is ``value-driven'', which instructs the behaviors of agents. The second is the agent’s autonomy in generating tasks for itself. 
The third is the generalization between different tasks, which requires the agent can transfer its abilities from one task to another through learning and adaptation. 
The mentioned features illustrate the portrait of high-level AI, and gradually become the Polaris of AGI development.

\subsection{The Shift of Cognition Paradigm}
The rise of high-level artificial intelligence will reshape the relationship between humans and AI agents. Since the research of XR usually focus on the human cognition in fully or partially immersive environments, the AGI will extend the cognition system from one to two centers, which challenges the conventional human-centric cognition paradigm.

\section{Human-Machine Interaction Frameworks}
Frameworks akin to symmetrical reality include ``digital twins'' and ``extended reality''. However, these frameworks elucidate issues from a perspective of world representation, overlooking the autonomous perception and interaction phenomena of AI entities. A fundamental distinction exists between these frameworks and the symmetrical reality. Symmetrical reality represents a novel human-machine interaction paradigm birthed from the amalgamation of virtual and physical worlds and adopts the dual-center architecture. We believe this fresh perspective will engender a plethora of innovative ideas for the human-machine interaction and virtual reality communities.

\begin{figure}[tb]
    \centering
    \includegraphics[width=\linewidth]{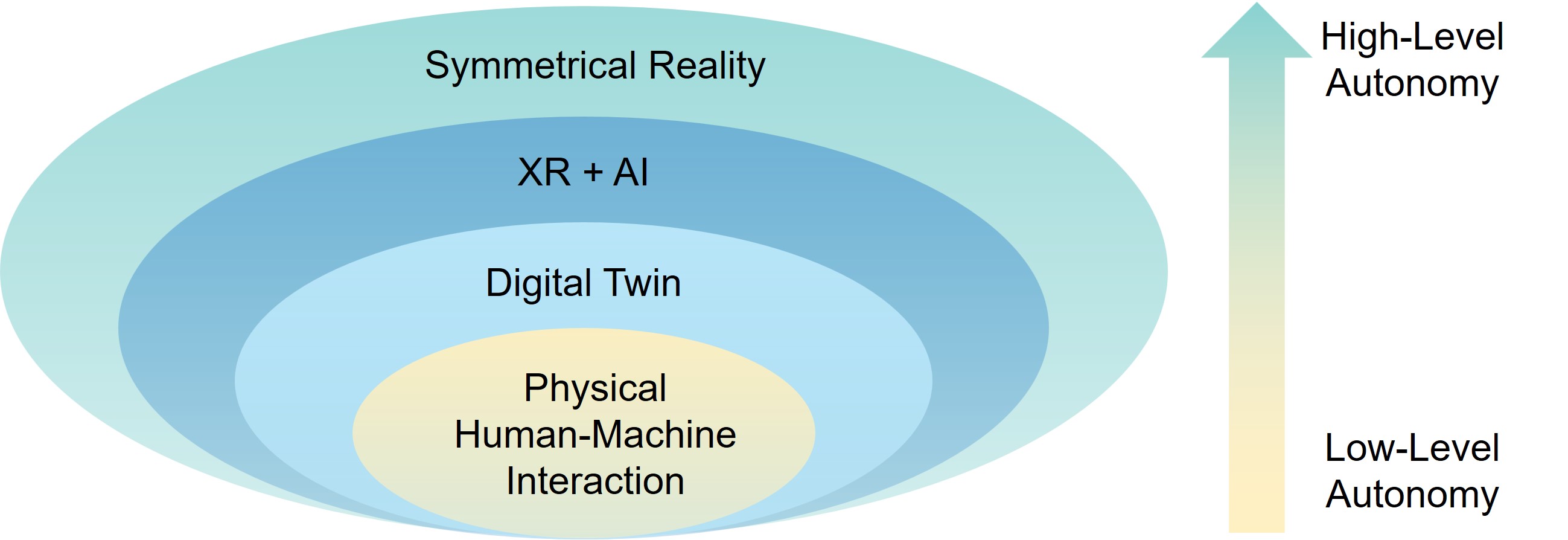} 
    \caption{Levels of symmetrical reality. The elliptic structure from small to large shows the inclusion relationship of different frameworks. The upgrading process from low-level to high-level autonomy promotes the expansion and evolution of the frameworks.}
    \label{fig:levels} 
\end{figure}

To deeply understand the relationship between SR and other frameworks, we use Fig.~\ref{fig:levels} to illustrate how these related concepts are interacting with each other.
\begin{itemize}
    \item \textbf{Level 1: Physical Human-Machine Interaction.}
    Humans and machines interact in the physical world, where the machine may only have low-level autonomy. For example, cell phones or personal computers belong to this category of machines. 
    \item \textbf{Level 2: Digital Twin.}
    Humans and machines interact in the physical world and simultaneously map information to the virtual world. This kind of structure adds the virtual world to the physical world, and also emphasizes the real-time linking between physical and virtual elements. 
    \item \textbf{Level 3: Combining Extended Reality (XR) and AI.}
    Both humans and machines are embodied in different spaces to achieve cross-reality interaction. Compared to level 2, this level focuses on the integration of XR and AI, which leads to more intelligent phenomena across physical and virtual worlds. 
    \item \textbf{Level 4: Symmetrical Reality.}
    Both humans and machines are embodied in different spaces to achieve cross-reality interaction and construct a dual-center architecture of virtual and real spaces. This level depends on the high-level autonomy of machines. Humans and machines are symmetrical regarding the cognition processes. 
\end{itemize}

\section{Conclusion and Future Direction}

Symmetrical reality, in theory, seeks to address the rapid advancements of AI entities that challenge conventional human-machine interaction paradigms. Technologically, the goal is to establish a supportive infrastructure for harmonious human-machine coexistence. 
As an emerging research field, this direction is still full of unknowns and challenges, and requires a more inclusive and developmental perspective to promote various research work, in order to contribute forward-looking solutions to the future society of human-machine coexistence.



\end{document}